# CLASSICAL BEHAVIOUR OF THE DIRAC BISPINOR

Sarah B. M. Bell,[1] John P. Cullerne[1] and Bernard M. Diaz[1, 2]

## ABSTRACT

It is usually supposed that the Dirac and radiation equations predict that the phase of a fermion will rotate through half the angle through which the fermion is rotated, which means, via the measured dynamical and geometrical phase factors, that the fermion must have a half-integral spin. We demonstrate that this is not the case and that the identical relativistic quantum mechanics can also be derived with the phase of the fermion rotating through the same angle as does the fermion itself. Under spatial rotation and Lorentz transformation the bispinor transforms as a four-vector like the potential and Dirac current. Previous attempts to provide this form of transformational behaviour have foundered because a satisfactory current could not be derived.(14)

---

[1] Department of Computer Science I.Q. Group, The University of Liverpool, Chadwick Building, Peach Street, Liverpool, L69 7ZF, United Kingdom.

[2] To whom correspondence should be addressed.

# 1. INTRODUCTION

## 1.1 Note on nomenclature

Matrices consisting of more than one row or column, quaternions, maps, and lifts are given boldface type. * signifies complex conjugation. $^T$ signifies transposition. $^\dagger$ signifies Hermitian conjugation. $^\ddagger$ signifies quaternion conjugation. A lowercase Latin subscript stands for 1, 2 or 3 and indicates the space axes. A lowercase Greek subscript stands for 0, 1, 2, or 3 and indicates the spacetime axes. $i$ is the square root $-1$. $\mathbf{i}_0 = 1$. $\mathbf{i}_1 = \mathbf{i}$, $\mathbf{i}_2 = \mathbf{j}$, $\mathbf{i}_3 = \mathbf{k}$ stand for the quaternion matrices, where $\mathbf{i}_r^2 = -1$ and $\mathbf{i}_1\mathbf{i}_2 = \mathbf{i}_3$, $\mathbf{i}_2\mathbf{i}_1 = -\mathbf{i}_3$ with cyclic variations. We have $\mathbf{i}_1{}^* = -\mathbf{i}_1$, $\mathbf{i}_2{}^* = \mathbf{i}_2$, $\mathbf{i}_3{}^* = -\mathbf{i}_3$, $\mathbf{i}_1^T = \mathbf{i}_1$, $\mathbf{i}_2^T = -\mathbf{i}_2$, $\mathbf{i}_3^T = \mathbf{i}_3$, $\mathbf{i}_r^\dagger = -\mathbf{i}_r$ and $\mathbf{i}_r^\ddagger = -\mathbf{i}_r$. Repetition of a subscript does not imply summation which is always signalled by the explicit use of $\sum$.

## 1.2 Background

We start with some definitions. Let Q be a quaternion[3, 5, 52] such that

$$\mathbf{Q} = q_0 + \mathbf{i}_1 q_1 + \mathbf{i}_2 q_2 + \mathbf{i}_3 q_3 \tag{1}$$

where the $q_\mu$ may be complex. We may associate a vector $Q$ with $\mathbf{Q}$, defining

$$Q = (q_0,\, q_1,\, q_2,\, q_3) \tag{2}$$

We will work with Euclidean spacetime,[58] in which the time axis is imaginary and the spatial axes real. Provided $q_0$ is imaginary and the $q_r$ are real, we may site $Q$ in Euclidean spacetime. This means that Lorentz transformations appear in the guise of rotations. By Lorentz transformations we will always mean proper and future-preserving Lorentz transformations in the language of Synge.[(61)] $Q$ is a *Euclidean four-vector* if it transforms under spatial rotation and Lorentz transformation like the Euclidean forms of the electromagnetic potential and Dirac current, preserving lengths and angles. These transformations may be effected by using quaternion multiplications acting on **Q**. There are many ways of doing this. The one we have described here is also discussed by, for example, Gough.[(36)] It is the method used by Edmonds.[(14)] If $Q$ is a four-vector, the transformational behaviour for **Q** for spatial rotations and Lorentz transformations, respectively, may be given by

$$\mathbf{Q}' = \mathbf{P}\mathbf{Q}\mathbf{P}^{\ddagger}, \qquad \mathbf{Q}' = \mathbf{P}\mathbf{Q}\mathbf{P} \tag{3}$$

for some suitable quaternion **P**. We may contrast this with the usually assumed half-angle behaviour of a bispinor $Q$ in which the transformational behaviour may be given by

$$\mathbf{Q}' = \mathbf{P}\mathbf{Q} \tag{4}$$

In this paper we derive a quaternion form of the Dirac equation, giving the behaviour of a fermion in an electromagnetic field, and the radiation equation, giving the electromagnetic field generated by a fermion. The quaternion Dirac equation has been studied before.[(1, 2, 6, 8-11, 13-28, 31-38, 41-44, 46, 47, 49-51, 53-57, 59-63)] Most of these tacitly or avowedly employ the usual half-angle behaviour of the bispinor, given in equation (4), under rotation or Lorentz transformation. This means that the phase of the spinor shows half-angle behaviour as well. There are three exceptions. Lambek,[(42)] Edmonds,[(19, 24)] and Edmonds[(14)] consider four-vector behaviour of the bispinor as given in equations (3), which implies whole-angle behaviour of the phase of the spinor. Lambek remarks that such behaviour leaves his version of the Dirac equation invariant but then considers it no further. Edmonds[(19)] finds satisfactory behaviour for his Dirac equation with no mass term and goes on to derive the free radiation equation by this route. Edmonds[(14)] shows that promoting the bispinor to four-vector behaviour with a variable mass term can leave his Dirac equation invariant. Here the spinor is a column vector whose entries are the quaternion bispinors. However, Edmonds[(14)] then shows that no satisfactory current can be derived from his Dirac equation in this form and concludes relativistic quantum mechanics cannot be rescued with this transformational behaviour.

Our new contribution uses *reflector* and *rotator matrices.* If $\mathbf{Q}$ and $\mathbf{U}$ are quaternions, we define *a reflector matrix* as having the form

$$\begin{pmatrix} 0 & \mathbf{Q} \\ \mathbf{U} & 0 \end{pmatrix}$$

while *a rotator matrix* has the form

$$\begin{pmatrix} \mathbf{Q} & 0 \\ 0 & \mathbf{U} \end{pmatrix}$$

We adopt a novel representation for the spinor as a reflector matrix whose quaternion elements are the bispinors. All the other variables in the Dirac equation are also represented as reflectors and the Dirac current, given by the trace of a rotator matrix, is also defined by reflector matrices. The transformational behaviour of reflector and rotator matrices can be found from the behaviour of their quaternion elements, for example that given in equations (3) and (4). Using the transformational behaviour of reflector and rotator matrices under spatial rotation, Lorentz transformation, parity change, time reversal and charge conjugation, we can then demonstrate not only that this form of the Dirac equation is invariant, but that it does have a satisfactory conserved current when both the bispinor and the mass term vary like a Euclidean four-vector. The current is identical to the original Dirac current and the eigenvalues identical to the original Dirac eigenvalues. In short the whole of relativistic quantum mechanics is recreated for four-vector classical behaviour of the Dirac bispinor and thus whole-angle behaviour of the phase of the spinor.

### 1.3 Outline

We discuss the replacement of contraction operations using the Pauli matrices with quaternion multiplications and additions. We derive a new version of the Dirac equation from the original in which all the variables are reflector matrices and we call the new version *the versatile Dirac equation*. We show that the two forms are completely equivalent in the sense that every solution of one is also a solution of the other with the same eigenvalues. We put the Dirac current in a compatible versatile form. We derive *the versatile radiation equation* from the original radiation equation.

We discuss the transformational properties of quaternions representing Euclidean four-vectors in Euclidean spacetime under spatial rotation and Lorentz transformation. We apply the results to reflector matrices. We demonstrate the invariance of the versatile Dirac equation under spatial rotation and Lorentz transformation for a range of possible angular behaviour of the phase of the spinor including Euclidean four-vector behaviour of the bispinor. For the same range of behaviour we demonstrate the invariance of the versatile Dirac equation under parity change, time reversal and charge conjugation. We show that the Euclidean Dirac current has Euclidean four-vector transformational behaviour under spatial rotation and Lorentz transformation under the same range of possible angular behaviour for the phase of the bispinor. This range includes Euclidean four-vector behaviour of the bispinor. We show that the current is conserved in all frames.

## 2. CONTRACTION OPERATIONS AND QUATERNION MULTIPLICATIONS

### 2.1 Contraction to a Vector

The essential elements of the lift and the map discussed in section 2 have appeared previously in the literature.[39, 40]

Let $\mathsf{H}^{\mathsf{C}}$ be the complexified algebra of quaternions $(q_\mu \mathbf{i}_\mu, q_\mu \in \mathsf{C})$ where $\mathsf{C}$ is the complex field. We define the $\mathsf{C}$-linear map

$$\mathbf{F}: \mathrm{H}^{\mathrm{C}} \rightarrow \mathrm{C}^{2}, \qquad \mathbf{Q} \rightarrow \mathbf{Q}\begin{pmatrix}1\\0\end{pmatrix}$$

where **Q** is a complex quaternion, and

$$\mathbf{i}_0 = \begin{pmatrix}1 & 0\\0 & 1\end{pmatrix}, \qquad \mathbf{i}_1 = \begin{pmatrix}0 & -i\\-i & 0\end{pmatrix},$$
$$\mathbf{i}_2 = \begin{pmatrix}0 & -1\\1 & 0\end{pmatrix}, \qquad \mathbf{i}_3 = \begin{pmatrix}-i & 0\\0 & i\end{pmatrix} \tag{5}$$

Then

$$\mathbf{F}(\mathbf{QU}) = \mathbf{QF}(\mathbf{U}) \tag{6}$$

where **U** is a complex quaternion, and the following further properties may be derived

$$\mathbf{F}(\mathbf{Q}(-\mathbf{i}_3)) = i\mathbf{F}(\mathbf{Q}), \qquad \mathbf{F}(\mathbf{Q}(1 + i\mathbf{i}_3)) = 2\mathbf{F}(\mathbf{Q}), \tag{7}$$

We note that the first of equations (7) provides the same answer as Rotelli[(54)] for translating the "*i*" of the complex representation into the quaternion representation.

We also define the lift **G**

$$\mathbf{G}: \mathrm{C}^{2} \rightarrow \mathrm{H}^{\mathrm{C}}, \qquad \mathbf{i}_\mu\begin{pmatrix}1\\0\end{pmatrix} \rightarrow \mathbf{i}_\mu$$

Then

$$\mathbf{F}^\circ\mathbf{G} = id_{\mathrm{C}^2} \tag{8}$$

Note that **G** is R-linear but not C-linear where R is the field of reals.

The following further properties may be derived for the lift **G**,

$$\mathbf{G}(\mathbf{i}_\mu \mathbf{F}(\mathbf{Q})) = \mathbf{i}_\mu \mathbf{Q}, \quad \mathbf{G}(i\mathbf{i}_\mu \mathbf{F}(\mathbf{Q})) = \mathbf{i}_\mu \mathbf{Q}(-\mathbf{i}_3) \tag{9}$$

## 2.2 Contraction to a Scalar

Let

$$\begin{aligned} \mathbf{U} &= u_0 + \mathbf{i}_1 u_1 + \mathbf{i}_2 u_2 + \mathbf{i}_3 u_3, \\ \mathbf{Q} &= q_0 + \mathbf{i}_1 q_1 + \mathbf{i}_2 q_2 + \mathbf{i}_3 q_3 \end{aligned} \tag{10}$$

where $q_\mu$ and $u_\mu$ may be complex. We call $u_0$, $u_1$, $u_2$ and $u_3$ *the components of* **U**. We define *the dot product of two quaternions*,

$$\mathbf{Q} \cdot \mathbf{U} = u_0 q_0 + u_1 q_1 + u_2 q_2 + u_3 q_3 \tag{11}$$

As may be easily seen by explicit enumeration,

$$\mathbf{Q} \cdot \mathbf{U} = (\mathbf{Q}^\ddagger \mathbf{U} + \mathbf{U}^\ddagger \mathbf{Q})/2 \tag{12}$$

where $^\ddagger$ signifies quaternion conjugation so that

$$\mathbf{U}^\ddagger = S - \mathbf{V} \tag{13}$$

where

$$\begin{aligned} S &= u_0, \quad \mathbf{V} = \mathbf{i}_1 u_1 + \mathbf{i}_2 u_2 + \mathbf{i}_3 u_3, \\ |\mathbf{U}| &= \mathbf{U}^\ddagger \mathbf{U}, \quad \mathbf{U}^{-1} = \mathbf{U}^\ddagger / |\mathbf{U}| \end{aligned} \tag{14}$$

We call S *the temporal part* of **U** and **V** *the spatial part.*

We then have for a further contraction operation,

$$\text{Real part of}\left\{[\mathbf{F}(\mathbf{Q})]^{\dagger}\mathbf{i}_r\mathbf{F}(\mathbf{U})\right\} = \mathbf{Q}\cdot(\mathbf{i}_r\mathbf{U}) \tag{15}$$

from equations (6) and (11) and where $^{\dagger}$ signifies Hermitian conjugation. Hence from equation (12),

$$\text{Real part of}\left\{[\mathbf{F}(\mathbf{Q})]^{\dagger}\mathbf{i}_r\mathbf{F}(\mathbf{U})\right\} = \left(\mathbf{Q}^{\ddagger}\mathbf{i}_r\mathbf{U} + \mathbf{U}^{\ddagger}\mathbf{i}_r^{\ddagger}\mathbf{Q}\right)/2 \tag{16}$$

We now consider a second contraction operation using a Pauli matrix, $\boldsymbol{\sigma}_r$.[7] Since $-i\boldsymbol{\sigma}_r$ is identically $\mathbf{i}_r$,

$$\text{Real part of}\left\{[\mathbf{F}(\mathbf{Q})]^{\dagger}\boldsymbol{\sigma}_r\mathbf{F}(\mathbf{U})\right\} = \mathbf{Q}\cdot\{\mathbf{i}_r\mathbf{U}(-\mathbf{i}_3)\} \tag{17}$$

from equations (6), (9) and (11). Therefore

$$\begin{aligned}\text{Real part of}&\left\{[\mathbf{F}(\mathbf{Q})]^{\dagger}\boldsymbol{\sigma}_r\mathbf{F}(\mathbf{U})\right\} \\ &= \left\{\mathbf{Q}^{\ddagger}\mathbf{i}_r\mathbf{U}(-\mathbf{i}_3) + (-\mathbf{i}_3)^{\ddagger}\mathbf{U}^{\ddagger}\mathbf{i}_r^{\ddagger}\mathbf{Q}\right\}/2\end{aligned} \tag{18}$$

from equation (12).

## 3. DERIVATION OF THE VERSATILE EQUATIONS

### 3.1 The Versatile Dirac Equation

We now turn the Dirac equation into the versatile Dirac equation. We also prove that the original and versatile versions are fully equivalent providing the fermion mass term remains a scalar.

The Dirac equation[7] is

$$i\left(\partial\Psi/\partial x_0^{\sim}\right)=\left\{\sum_r\left(-i\alpha_r\left(\partial/\partial x_r\right)-\alpha_r A_r^{\sim}\right)+A_0^{\sim}+m^{\sim}\beta\right\}\Psi \tag{19}$$

where $\Psi$ is the wave function and a column matrix, $x_0^{\sim}$ the time co-ordinate, $A^{\sim}$ the potential Minkowski four-vector in Minkowski spacetime,

$$A^{\sim}=\left(A_0^{\sim},A_1^{\sim},A_2^{\sim},A_3^{\sim}\right) \tag{20}$$

and $m^{\sim}$ the rest mass of the fermion. We define the bispinors by

$$\Psi=\left(\psi_1,\psi_2\right)^T \tag{21}$$

and the matrices by

$$\alpha_r=\begin{pmatrix}0 & \sigma_r\\ \sigma_r & 0\end{pmatrix},\qquad \beta=\begin{pmatrix}1 & 0\\ 0 & -1\end{pmatrix} \tag{22}$$

We now translate the original Dirac equation (19) into the versatile version. The same process may be followed to obtain the versatile version of any operator. We start by translating to Euclidean spacetime by making the following changes of variable,

$$m=m^{\sim}/i \tag{23}$$

$$A_r=A_r^{\sim},\qquad A_0=A_0^{\sim}/i \tag{24}$$

$$\left(\phi_1,\phi_2\right)^T=\mathbf{V}\left(\psi_1,\psi_2\right)^T \tag{25}$$

and

$$\mathbf{V} = \begin{pmatrix} 1 & 1 \\ i & -i \end{pmatrix} \tag{26}$$

The new wave function is

$$\Phi = (\phi_1, \phi_2)^T \tag{27}$$

We define *the bijection* **H** between vectors in $\mathbb{C}^4$ and quaternions

$$(v_0, v_1, v_2, v_3) \rightarrow (v_0 + \mathbf{i}_1 v_1 + \mathbf{i}_2 v_2 + \mathbf{i}_3 v_3)$$

where the $v_\mu$ may be complex. We call the left-hand-side *the vector associated with the quaternion* on the right-hand side. We then have

$$\mathbf{D} = \mathbf{H}\{(i\,\partial/\partial x_0^{\sim}\,, \partial/\partial x_1\,, \partial/\partial x_2\,, \partial/\partial x_3)\} \tag{28}$$

$$\mathbf{A} = \mathbf{H}\{(A_0, A_1, A_2, A_3)\} \tag{29}$$

and equation (19) becomes

$$(\mathbf{D} - i\mathbf{A})^{\ddagger}\phi_1 = m\phi_2, \qquad (\mathbf{D} - i\mathbf{A})\phi_2 = -m\phi_1 \tag{30}$$

We may now use the lift **G** of section 2.1,

$$\begin{gathered} \mathbf{G}\{(\mathbf{D} - i\mathbf{A})^{\ddagger}\phi_1 - m\phi_2\} = 0, \\ \mathbf{G}\{(\mathbf{D} - i\mathbf{A})\phi_2 + m\phi_1\} = 0 \end{gathered} \tag{31}$$

From the lift **G** and equations (9), (23) and (24) we obtain for the first of equations (31)

$$\mathbf{G}\left(\partial\phi_1/\partial x_0^{\sim}\right)(-\mathbf{i}_3)+\sum_r\left(-\mathbf{i}_r\mathbf{G}\left(\partial\phi_1/\partial x_r\right)\right)-A_0^{\sim}\mathbf{G}(\phi_1)$$

$$+\sum_r\left(\mathbf{i}_r A_r \mathbf{G}(\phi_1)(-\mathbf{i}_3)\right)+m^{\sim}\mathbf{G}(\phi_2)(-\mathbf{i}_3)=0 \qquad (32)$$

and a similar equation for the second of equations (31). We postmultiply both the equations so derived by $(1 + i\mathbf{i}_3)$. We obtain, again using equations (23) and (24),

$$(\mathbf{D}-i\mathbf{A})^{\ddagger}\phi_1=\phi_2\mathbf{M}, \quad (\mathbf{D}-i\mathbf{A})\phi_2=-\phi_1\mathbf{M}^{\ddagger} \qquad (33)$$

where

$$\mathbf{M}=\mathbf{M}^{\ddagger}=m \qquad (34)$$

and we choose to write **M** in deference to properties we discuss in section 5.1, and

$$\begin{aligned}\phi_1^{\S}&=\mathbf{G}(\phi_1), & \phi_2^{\S}&=\mathbf{G}(\phi_2),\\ \phi_1&=\phi_1^{\S}(1+i\mathbf{i}_3), & \phi_2&=\phi_2^{\S}(1+i\mathbf{i}_3)\end{aligned} \qquad (35)$$

Equation (33) is the versatile Dirac equation in the bispinors and equation (35) defines the versatile bispinors. We have just proved that every solution of the original Dirac equation (19) corresponds to a solution of the versatile Dirac equation (33), with the same eigenvalues. We see that the bispinors $\phi_1$ and $\phi_2$ satisfy the first part of Rastall's[(53)] definition of a spinor (his terminology) in that the bispinors (our terminology) are the left ideal generated by the idempotent $(1 + i\mathbf{i}_3)/2$. We will see that our bispinors satisfy the second part of his definition in section 4.

We may map equation (33) with **F**,

$$\mathbf{F}\left\{(\mathbf{D}-i\mathbf{A})^{\ddagger}\phi_1-\mathbf{M}\phi_2\right\}=0, \qquad (36)$$
$$\mathbf{F}\left\{(\mathbf{D}-i\mathbf{A})\phi_2+\mathbf{M}^{\ddagger}\phi_1\right\}=0$$

and from the definition of map **F** and equations (6), (7), or (8) we obtain equation (30). We have now proved that every solution of the versatile Dirac equation (33) corresponds to a solution of the original Dirac equation (19). Thus, while **M** remains a scalar equal to *m*, the versatile Dirac equation is simply another form of the original equation.

We recall our definition of *reflector matrices* as

$$\underline{\mathbf{U}}=\underline{\mathbf{U}}(\mathbf{Q},\mathbf{U})=\begin{pmatrix}0 & \mathbf{Q}\\ \mathbf{U} & 0\end{pmatrix} \qquad (37)$$

where **Q** and **U** are complex quaternions or complex numbers. The versatile Dirac equation in the bispinors then becomes the versatile Dirac equation in the spinor, $\underline{\mathbf{\Phi}}(\phi_1,\phi_2)$,

$$\left(\underline{\mathbf{D}}(\mathbf{D},\mathbf{D}^{\ddagger})-i\,\underline{\mathbf{A}}(\mathbf{A},\mathbf{A}^{\ddagger})\right)\underline{\mathbf{\Phi}}(\phi_1,\phi_2)=\underline{\mathbf{\Phi}}(\phi_1,\phi_2)\underline{\mathbf{M}}(\mathbf{M},-\mathbf{M}^{\ddagger})$$

or

$$(\underline{\mathbf{D}}-i\,\underline{\mathbf{A}})\,\underline{\mathbf{\Phi}}=\underline{\mathbf{\Phi}}\,\underline{\mathbf{M}} \qquad (38)$$

## 3.2 The Versatile Radiation Equation

A conserved current, the Dirac current, may be derived from the original Dirac equation, equation (19), with zero potential $A_\mu$.[(7)] We want this current in reflector form so that its behaviour under transformation can be compared with that of the versatile Dirac equation. The conserved current from the original Dirac equation is

$$J_0^{\S\sim} = \Psi^\dagger \Psi, \qquad J_r^{\S\sim} = \Psi^\dagger \boldsymbol{\alpha}_r \Psi \tag{39}$$

Translating this into our new variables using equations (25) and (26) and transforming to Euclidean spacetime with

$$J_0^{\S} = J_0^{\S\sim}/i, \qquad J_r^{\S} = J_r^{\S\sim} \tag{40}$$

equation (39) becomes

$$J_0^{\S} = -i\left(\phi_1^\dagger \phi_1 + \phi_2^\dagger \phi_2\right)/2, \qquad J_r^{\S} = \left(\phi_1^\dagger \boldsymbol{\sigma}_r \phi_1 - \phi_2^\dagger \boldsymbol{\sigma}_r \phi_2\right)/2 \tag{41}$$

Changing variables in equations (41) using equations (12) and (35) for the first and equations (18) and (35) for the second,

$$J_0^{\S} = -i\left(\boldsymbol{\phi}_1^{\S\ddagger}\boldsymbol{\phi}_1^{\S} + \boldsymbol{\phi}_2^{\S\ddagger}\boldsymbol{\phi}_2^{\S}\right)/2$$

$$J_r^{\S} = \left(\boldsymbol{\phi}_1^{\S\ddagger}\mathbf{i}_r^{\ddagger}\boldsymbol{\phi}_1^{\S}\mathbf{i}_3 + \mathbf{i}_3\boldsymbol{\phi}_1^{\S\ddagger}\mathbf{i}_r^{\ddagger}\boldsymbol{\phi}_1^{\S} + \boldsymbol{\phi}_2^{\S\ddagger}\mathbf{i}_r\boldsymbol{\phi}_2^{\S}\mathbf{i}_3 + \mathbf{i}_3\boldsymbol{\phi}_2^{\S\ddagger}\mathbf{i}_r\boldsymbol{\phi}_2^{\S}\right)/4 \tag{42}$$

where $J_0^{\S}$ and $J_r^{\S}$ are quaternions with only a temporal part. After the lift $\mathbf{G}$ the components of $\boldsymbol{\phi}_1^{\S}$ and $\boldsymbol{\phi}_2^{\S}$ are real. Hence we may set

$$\boldsymbol{\phi}_1^{\S\ddagger} = \boldsymbol{\phi}_1^{\S\dagger}, \qquad \boldsymbol{\phi}_2^{\S\ddagger} = \boldsymbol{\phi}_2^{\S\dagger} \tag{43}$$

in equation (42) as may be easily checked. We also pre- and post-multiply equation (42) by $\left(1 + i\mathbf{i}_3\right)$ and obtain from equation (35)

$$J_{\mu}^{\S} = t:\left\{\mathbf{k}_{\mu}\left(\phi_1^{\dagger}\mathbf{i}_{\mu}^{\ddagger}\phi_1 + \phi_2^{\dagger}\mathbf{i}_{\mu}\phi_2\right)\right\} \tag{44}$$

where t: stands for *the temporal part of* and $\mathbf{k}_{\mu}$ is $-i/4$.

Let

$$\mathbf{J}_{\mu}^{1} = \mathbf{k}_{\mu}\phi_2^{\dagger}\mathbf{i}_{\mu}\phi_2, \qquad \mathbf{J}_{\mu}^{2} = \mathbf{k}_{\mu}^{\ddagger}\phi_1^{\dagger}\mathbf{i}_{\mu}^{\ddagger}\phi_1 \tag{45}$$

where we note that $\mathbf{k}_{\mu}$ and $\mathbf{k}_{\mu}^{\ddagger}$ are identical. We recall our definition of *rotator matrices* for any $\mathbf{U}|$ as

$$\mathbf{U}| = \mathbf{U}|\left(\mathbf{Q},\mathbf{U}\right) = \begin{pmatrix}\mathbf{Q} & 0 \\ 0 & \mathbf{U}\end{pmatrix} \tag{46}$$

where **Q** and **U** are quaternions or complex numbers. We define

$$t:\underline{\mathbf{U}} = \underline{\mathbf{U}}\left(t:\mathbf{Q},\, t:\mathbf{U}\right), \qquad t:\mathbf{U}| = \mathbf{U}|\left(t:\mathbf{Q},\, t:\mathbf{U}\right) \tag{47}$$

and that $\mathbf{U}|^{\ddagger}$ or $\underline{\mathbf{U}}^{\ddagger}$ is formed by taking the quaternion conjugate of any quaternion elements.

Let

$$\begin{aligned}
\mathbf{J}_{\mu}| &= \mathbf{J}_{\mu}|\left(\mathbf{J}_{\mu}^{1},\, \mathbf{J}_{\mu}^{2}\right), \qquad & \underline{\mathbf{\Phi}}_{S} &= \underline{\mathbf{\Phi}}_{S}\left(\phi_1^{\dagger},\, \phi_2^{\dagger}\right), \\
\underline{\mathbf{I}}_{\mu} &= \underline{\mathbf{I}}_{\mu}\left(\mathbf{i}_{\mu},\mathbf{i}_{\mu}^{\ddagger}\right), \qquad & \underline{\mathbf{K}}_{\mu} &= \underline{\mathbf{K}}_{\mu}\left(\mathbf{k}_{\mu},\, \mathbf{k}_{\mu}^{\ddagger}\right)
\end{aligned} \tag{48}$$

and we see from equation (45) that

$$\mathbf{J}_{\mu}| = \underline{\mathbf{K}}_{\mu}\underline{\mathbf{\Phi}}_{S}\,\underline{\mathbf{I}}_{\mu}\underline{\mathbf{\Phi}} \tag{49}$$

giving

$$\mathbf{J}^{\S}_{\mu} = t:\{\mathrm{Trace}(\mathbf{J}_{\mu}|)\} \tag{50}$$

from equation (44). Equations (49) and (50) define the current in versatile form.

We wish to relate the Dirac current to the potential it generates. We would like the versatile radiation equation in rotator and reflector form so that its behaviour under transformation can be compared with that of the versatile Dirac equation. Since $\mathbf{DD}^{\ddagger}$ and $\mathbf{D}^{\ddagger}\mathbf{D}$ are both the d'Alembertian, in the Lorentz gauge,[58] we have

$$\mathbf{DD}^{\ddagger} A_{\mu}\mathbf{i}_{\mu} = J^{\S}_{\mu}\mathbf{i}_{\mu}, \qquad \mathbf{D}^{\ddagger}\mathbf{D}A_{\mu}\mathbf{i}^{\ddagger}_{\mu} = J^{\S}_{\mu}\mathbf{i}^{\ddagger}_{\mu} \tag{51}$$

since these equations are equally true in Euclidean spacetime. We set

$$\mathbf{J} = \sum_{\mu} J^{\S}_{\mu}\mathbf{i}_{\mu} \tag{52}$$

Using equations (29) we may write equations (51) in reflector form

$$\underline{\mathbf{D}}\underline{\mathbf{D}}\underline{\mathbf{A}} = \underline{\mathbf{J}} \tag{53}$$

where

$$\underline{\mathbf{J}} = \underline{\mathbf{J}}(\mathbf{J}, \mathbf{J}^{\ddagger}) \tag{54}$$

Equation (53) is the versatile radiation equation.

# 4. ROTATION AND LORENTZ TRANSFORMATION OF REFLECTORS AND ROTATORS

## 4.1 Rotation and Lorentz Transformation in $\mathsf{R}^4$

If we may associate our Euclidean four-vectors with a quaternion using the map **H,** we may describe their spatial rotation and Lorentz transformation in two ways. In the first we transform the components of the quaternions, identically the elements of the Euclidean four-vectors, using a suitable spatial rotation or boost matrix and leave the quaternion matrices, $\mathbf{i}_\mu$, constant. In the second we leave the components of the quaternion constant and transform the quaternion matrices, $\mathbf{i}_\mu$, using quaternion multiplication. Dirac[7] chose the former course, while we will always choose the latter in this paper.

Therefore, we seek the geometry of quaternion and hence reflector and rotator multiplications, so that we may use them to describe spatial rotations and Lorentz transformations of Euclidean four-vectors and bispinors. This has been studied before. There is more than one way to do it.[12, 45, 50, 53, 61] Our method is the same as that used by Edmonds[14] and described by Gough,[36] among others, although we describe spatial rotations and Lorentz transformations separately for greater clarity. For a transparent proof that this is so see Synge.[61] Those who prefer a group theoretic discussion and less geometry may consult MacFarlane.[48] We start a general introduction by considering real quaternions in $\mathsf{R}^4$. We will call the co-ordinates of points in this space *four-vectors.*

Let **R** be a quaternion with real components and unit modulus with

$$\mathbf{R} = r_0 + \mathbf{i}_1 r_1 + \mathbf{i}_2 r_2 + \mathbf{i}_3 r_3 \tag{55}$$

and

$$\begin{aligned} S_R &= \mathbf{H}^{-1}(r_0) = (r_0, 0, 0, 0), \\ V_\mathrm{R} &= \mathbf{H}^{-1}(\mathbf{i}_1 r_1 + \mathbf{i}_2 r_2 + \mathbf{i}_3 r_3) = (0, r_1, r_2, \mathrm{r}_3) \end{aligned} \tag{56}$$

We call the plane containing $S_R$ and $V_R$ *the temporal plane* and the plane everywhere at right angles to it *the spatial plane*. We define

$$\tan(\theta/2) = \sqrt{\left(r_1^2 + r_2^2 + r_3^2\right)}\Big/ r_0 \tag{57}$$

Let **Q** be a quaternion with real components. Table I gives the angular behaviour of $\mathbf{H}^{-1}(\mathbf{Q})$ under quaternion multiplication by **R.** The first column defines the type of multiplication and the second two columns define the angle, $\xi$, through which **Q** turns to become $\mathbf{Q}'$. The first column gives the angle, $\xi_s$, in the spatial plane and the second the angle, $\xi_t$, in the temporal plane.

**Table I.** The Rotations Effected by Quaternion Multiplication

| Expression, $\mathbf{Q}'=$ | Spatial plane $\xi_s$ | Temporal plane $\xi_t$ |
|---|---|---|
| $\mathbf{RQ}$ | $+\xi/2$ | $+\xi/2$ |
| $\mathbf{QR}$ | $-\xi/2$ | $+\xi/2$ |
| $\mathbf{R}^{\ddagger}\mathbf{Q}$ | $-\xi/2$ | $-\xi/2$ |
| $\mathbf{QR}^{\ddagger}$ | $+\xi/2$ | $-\xi/2$ |
| $\mathbf{RQR}$ | 0 | $+\xi$ |
| $\mathbf{RQR}^{\ddagger}$ | $+\xi$ | 0 |
| $\mathbf{R}^{\ddagger}\mathbf{QR}$ | $-\xi$ | 0 |
| $\mathbf{R}^{\ddagger}\mathbf{QR}^{\ddagger}$ | 0 | $-\xi$ |

We call a rotation in the spatial plane *a spatial rotation* and a rotation in the temporal plane *a temporal rotation.*

We now develop the properties of the reflector matrices under rotation. We consider the quaternion equation

$$\mathbf{QP}^{\ddagger}=\mathbf{PW}^{\ddagger} \tag{58}$$

We may provide the four-vectors $\mathbf{H}^{-1}(\mathbf{Q})$, $\mathbf{H}^{-1}(\mathbf{P})$ and $\mathbf{H}^{-1}(\mathbf{W})$ with a spatial rotation by pre- and post-multiplication of the quaternions by a suitable $\mathbf{R}$ and $\mathbf{R}^{\ddagger}$. This transformation is a similarity transformation that leaves equation (58) invariant. However, if we wish to give the variables a temporal rotation, we must pre- and post-multiply $\mathbf{Q}$ and $\mathbf{P}$ by $\mathbf{R}$ while $\mathbf{P}^{\ddagger}$ and $\mathbf{W}^{\ddagger}$ must be pre- and post-multiplied by $\mathbf{R}^{\ddagger}$ to preserve the relation of quaternion conjugation. Neither of these is a similarity transformation although equation (58) remains invariant. However, if we take the quaternion conjugate of equation (58) and re-arrange we obtain a second form of the equation

$$\mathbf{Q}^{\ddagger}\mathbf{P} = \mathbf{P}^{\ddagger}\mathbf{W} \tag{59}$$

We may write equations (58) and (59) as

$$\underline{\mathbf{Q}}\left(\mathbf{Q},\mathbf{Q}^{\ddagger}\right)\underline{\mathbf{P}}\left(\mathbf{P},\mathbf{P}^{\ddagger}\right) = \underline{\mathbf{P}}\left(\mathbf{P},\mathbf{P}^{\ddagger}\right)\underline{\mathbf{W}}\left(\mathbf{W},\mathbf{W}^{\ddagger}\right) \tag{60}$$

Equation (60) implies no more than equation (58) but we may now provide a temporal rotation which leaves equation (60) invariant and is a similarity transformation, as follows. We set

$$\mathbf{R}_S = \mathbf{R}_T = \mathbf{R} \tag{61}$$

$$\mathbf{R}_S| = \mathbf{R}_S|\left(\mathbf{R}_S,\mathbf{R}_S\right) \tag{62}$$

$$\mathbf{R}_T| = \mathbf{R}_T|\left(\mathbf{R}_T,\mathbf{R}_T^{\ddagger}\right) \tag{63}$$

and we may effect a spatial rotation of $\mathbf{H}^{-1}(\mathbf{Q})$ by

$$\underline{\mathbf{Q}'} = \mathbf{R}_S|\,\underline{\mathbf{Q}}\mathbf{R}_S|^{\ddagger} \tag{64}$$

and a temporal rotation by

$$\underline{\mathbf{Q}}' = \mathbf{R}_T| \, \underline{\mathbf{Q}} \mathbf{R}_T|^{\ddagger} \tag{65}$$

and similarly for **P** and **W**.

## 4.2 Generalisation and Further Properties

We may broaden our remarks. We may include

$$\underline{\mathbf{Q}} = \underline{\mathbf{Q}}(\mathbf{Q}, \mathbf{U}) \tag{66}$$

$$\mathbf{Q}| = \mathbf{Q}|(\mathbf{Q}, \mathbf{U}) \tag{67}$$

where **Q** and **U** are distinct, providing that **U** rotates in the same sense as $\mathbf{Q}^{\ddagger}$ under a temporal rotation. This gives us for a spatial rotation

$$\underline{\mathbf{Q}}'(\mathbf{Q}', \mathbf{U}') = \mathbf{R}_S| \, \underline{\mathbf{Q}}(\mathbf{Q}, \mathbf{U}) \mathbf{R}_S|^{\ddagger} \tag{68}$$

and for a temporal rotation

$$\underline{\mathbf{Q}}'(\mathbf{Q}', \mathbf{U}') = \mathbf{R}_T| \, \underline{\mathbf{Q}}(\mathbf{Q}, \mathbf{U}) \mathbf{R}_T|^{\ddagger} \tag{69}$$

Further, all the reflectors in equation (60) may be distinct. Thus equations consisting of reflector multiplications are invariant under any rotation in $\mathrm{R}^4$ up to a similarity transformation.

Suppose we have an equation in which the product of an even number of reflector matrices, for example, $\underline{\mathbf{P}}(\mathbf{P}, \mathbf{S})$, is set equal to a rotator matrix, $\mathbf{Q}|(\mathbf{Q}, \mathbf{U})$. For a spatial rotation of the four-vectors associated with the quaternions in the $\underline{\mathbf{P}}(\mathbf{P}, \mathbf{S})$, we have for the rotator

$$\mathbf{Q}|' = \mathbf{R}_S| \, \mathbf{Q}| \, \mathbf{R}_S|^{\ddagger} \tag{70}$$

which provides a spatial rotation of **Q** and **U**. For a temporal rotation of the four-vectors associated with quaternions in the $\underline{\mathbf{P}}(\mathbf{P},\mathbf{S})$, we have for the rotator

$$\mathbf{Q}|' = \mathbf{R}_T| \, \mathbf{Q}| \, \mathbf{R}_T|^{\ddagger} \tag{71}$$

which also provides a spatial rotation of **Q** and **U**. Thus the temporal components of the quaternions associated with the rotator are constant under any type of rotation.

We may also derive a similar relation by considering the trace of rotator matrices. The trace is always zero for the product of an odd number of reflectors and so yields little of interest for equations with an odd number of reflectors either side. However, an even number of reflectors combine to form a rotator matrix and the trace is not in general zero. Since both spatial and temporal rotations of the four-vectors associated with the quaternions in the $\underline{\mathbf{P}}(\mathbf{P},\mathbf{S})$ are a similarity transformation for the equation we are guaranteed that this trace is invariant under any type of rotation. Inspection of the quaternion matrices of equation (5) tells us that the trace is formed from the temporal parts of the participating quaternions.

### 4.3 Rotations in Euclidean Spacetime

We may further broaden our remarks by placing the four-vector associated with our quaternion **Q** or **U** of the last section, in Euclidean spacetime as a Euclidean four-vector. We will continue to call the new quaternions **Q** and **U**, but although we have real spatial components from now on we have an imaginary temporal component. Spatial rotations are not affected, and we have

$$\underline{\mathbf{Q}}'(\mathbf{Q}',\mathbf{U}') = \mathbf{R}_S| \, \underline{\mathbf{Q}}(\mathbf{Q},\mathbf{U}) \mathbf{R}_S|^{\ddagger} \tag{72}$$

where $\mathbf{R}_S|$ is defined in (62) and $\mathbf{R}_S$ has real components. However, if we want $\mathbf{H}^{-1}(\mathbf{Q})$ to remain within Euclidean spacetime for a temporal rotation, we must revise the $\mathbf{R}_T$ in equation (63).

We do this by first considering a rotation matrix acting on the components of a quaternion and then translating this into quaternion multiplications. We consider only a rotation in the plane of the temporal and first spatial axes since a general temporal rotation may be built from this and spatial rotations. Let

$$\mathbf{Q} = q_0 + \mathbf{i}_1 q_1 + \mathbf{i}_2 q_2 + \mathbf{i}_3 q_3 \tag{73}$$

We may write the required transformation with a rotation matrix,

$$\begin{pmatrix} q_1' \\ q_0' \end{pmatrix} = \begin{pmatrix} \cos\theta & -\sin\theta \\ \sin\theta & \cos\theta \end{pmatrix} \begin{pmatrix} q_1 \\ q_0 \end{pmatrix} \tag{74}$$

and set

$$q_1^{\sim} = q_1, \quad q_1'^{\sim} = q_1', \quad q_0^{\sim} = iq_0, \quad q_0'^{\sim} = iq_0' \tag{75}$$

where the $q_\mu$ superscripted by $^{\sim}$ are real. Then equation (74) becomes

$$\begin{pmatrix} q_1'^{\sim} \\ q_0'^{\sim} \end{pmatrix} = \begin{pmatrix} \cosh\theta'' & \sinh\theta'' \\ \sinh\theta'' & \cosh\theta'' \end{pmatrix} \begin{pmatrix} q_1^{\sim} \\ q_0^{\sim} \end{pmatrix} \tag{76}$$

with

$$\cos\theta = \cosh\theta'', \qquad \sin\theta = \text{-i}(\sinh\theta''), \qquad i\theta = \theta'' \tag{77}$$

where $\theta''$ is real and we recognise the familiar form of a Lorentz transformation in equation (76).[29, 30]

We may now translate equation (74) into a relation using reflectors and rotators using the method set out in section 4.1. We set

$$\mathbf{R}_T = \cos(\theta/2) + \mathbf{i}_1 \sin(\theta/2) \tag{78}$$

where $\sin(\theta/2)$ is imaginary. Provided that $\mathbf{H}^{-1}(\mathbf{R}_T)$ is not on the light cone $\mathbf{R}_T$ has an inverse and we may form the rotator

$$\mathbf{R}_T| = \mathbf{R}_T|\left(\mathbf{R}_T, \mathbf{R}_T^{\ddagger}\right) \tag{79}$$

We may then provide the required boost by

$$\underline{\mathbf{Q}}' = \mathbf{R}_T|\underline{\mathbf{Q}}\mathbf{R}_T|^{\ddagger} \tag{80}$$

From now on we shall assume that, while $\mathbf{R}_S$ has real components, $\mathbf{R}_T$ has a real temporal component but imaginary spatial components.

We have now shown that all spatial rotations and Lorentz transformations of any Euclidean four-vectors associated with $\underline{\mathbf{Q}}$ can be effected by suitably chosen $\mathbf{R}_S|$ and $\mathbf{R}_T|$.

## 5. PHYSICAL PROPERTIES OF THE DIRAC AND RADIATION EQUATIONS

### 5.1 The Versatile Dirac Equation

We discuss the behaviour of the Dirac equation under spatial and temporal rotation in Euclidean spacetime. We consider the variables which always behave like Euclidean four-vectors first. From equations (72) and (80) we have

$$\underline{\mathbf{D}}' = \mathbf{R}| \, \underline{\mathbf{D}} \mathbf{R}|^{\ddagger} \tag{81}$$

$$\underline{\mathbf{A}}' = \mathbf{R}| \, \underline{\mathbf{A}} \, \mathbf{R}|^{\ddagger} \tag{82}$$

where $\mathbf{R}|$ may represent a spatial rotation,

$$\mathbf{R}| = \mathbf{R}_S| \left( \mathbf{R}_S, \mathbf{R}_S \right) \tag{83}$$

or a Lorentz transformation,

$$\mathbf{R}| = \mathbf{R}_T| \left( \mathbf{R}_T, \mathbf{R}_T^{\ddagger} \right) \tag{84}$$

Next we consider $\underline{\mathbf{M}}$, $\underline{\mathbf{\Phi}}$ and the transformation formulae

$$\underline{\mathbf{M}}' = \mathbf{R}|^{n} \underline{\mathbf{M}} \mathbf{R}|^{\ddagger n} \tag{85}$$

$$\underline{\mathbf{\Phi}}' = \mathbf{R}| \, \underline{\mathbf{\Phi}} \, \mathbf{R}|^{\ddagger n} \tag{86}$$

With the understanding that a matrix to the power zero is the unit matrix, $n = 0$ gives the usually quoted transformational properties of the Dirac equation[7] and we see that our bispinor satisfies the second of Rastall's[53] conditions. However, any integer value of *n* will leave the versatile Dirac equation in an invariant form, as may easily be checked.

If we apply a temporal rotation with $n \neq 0$ to the versatile Dirac equation (38), we may not subsequently map it back to the original Dirac equation (19). This is because **M** has become a quaternion with a nonzero spatial component or nonzero components and we may not shift the quaternions $\phi_1$ or $\phi_2$ to the right hand side of **M** in going from equation (33) to equation (36). At this point we have generalised the original Dirac equation. However, since the transformational properties of **D** and **A** have not been altered, the eigenvalues associated with the versatile Dirac equation for $n \neq 0$ continue the same as those of the original Dirac equation where $n = 0$ transformational behaviour is assumed.

For $n = 1,$ the vector associated with the mass term **M** has the transformational behaviour of a Euclidean four-vector and so do the vectors associated with the bispinors $\phi_1$ or $\phi_2^\ddagger,$ justifying the title of our paper. We note, however, that the position of the versatile bispinor depends on whether we map spin up or spin down to the space addressed by the complexified axes $x_1$ and $x_2$ in $\mathsf{C}^4$. We may cater for this by adding a second choice of lift

$$\mathbf{L} : \mathrm{C}^2 \to \mathrm{H}^{\mathrm{C}}, \qquad \mathbf{i}_\mu \begin{pmatrix} 0 \\ 1 \end{pmatrix} \to \mathbf{i}_\mu$$

accompanied by a map **N,**

$$\mathbf{N} : \mathrm{H}^{\mathrm{C}} \to \mathrm{C}^2, \qquad \mathbf{Q} \to \mathbf{Q}\begin{pmatrix} 0 \\ 1 \end{pmatrix}$$

where **Q** is any complex quaternion.

If we regard **M** as a variable rather than a constant, that is, for $n \neq 0$, we may still recover invariance under a parity change or time reversal. We have

$$\underline{\mathbf{D}}'' = \underline{\mathbf{B}}\,\underline{\mathbf{D}}\,\underline{\mathbf{B}}^{\ddagger} \tag{87}$$

and similarly for $\underline{\mathbf{A}}$,

$$\underline{\mathbf{\Phi}}'' = \underline{\mathbf{B}}\,\underline{\mathbf{\Phi}}\,\underline{\mathbf{E}}^{\ddagger} \tag{88}$$

$$\underline{\mathbf{M}}'' = \underline{\mathbf{E}}\,\underline{\mathbf{M}}\,\underline{\mathbf{E}}^{\ddagger} \tag{89}$$

where

$$\underline{\mathbf{B}} = \underline{\mathbf{C}}_{+}(1,1), \qquad \underline{\mathbf{E}} = \underline{\mathbf{C}}_{-}(-1,1) \tag{90}$$

for a parity change, and

$$\underline{\mathbf{B}} = \underline{\mathbf{C}}_{-}(-1,1), \qquad \underline{\mathbf{E}} = \underline{\mathbf{C}}_{+}(1,1) \tag{91}$$

for time reversal. Either choice leaves equation (38) invariant.

For charge conjugation we take the complex conjugate of equation (38), giving

$$\left(\underline{\mathbf{D}}^* + i\,\underline{\mathbf{A}}^*\right)\underline{\mathbf{\Phi}}^* = \underline{\mathbf{\Phi}}^*\,\underline{\mathbf{M}}^* \tag{92}$$

where $\underline{\mathbf{D}}^*$ means that the complex conjugate is taken of the elements of $\underline{\mathbf{D}}$ etc. Let

$$\mathbb{T} = \mathbb{T}\left(\mathbf{i}_2, \mathbf{i}_2\right) \tag{93}$$

If we then transform the variables in equation (92) by

$$\underline{\mathbf{D}}'' = \mathbb{T}\,\underline{\mathbf{C}}_-\underline{\mathbf{D}}^*\underline{\mathbf{C}}_-^\dagger\mathbb{T}^\dagger, \qquad \underline{\mathbf{A}}'' = \mathbb{T}\,\underline{\mathbf{C}}_-\underline{\mathbf{A}}^*\underline{\mathbf{C}}_-^\dagger\mathbb{T}^\dagger,$$

$$\underline{\mathbf{M}}'' = \mathbb{T}\,\underline{\mathbf{C}}_+\underline{\mathbf{M}}^*\underline{\mathbf{C}}_+^\dagger\mathbb{T}^\dagger, \qquad \underline{\mathbf{\Phi}}'' = \mathbb{T}\,\underline{\mathbf{C}}_-\underline{\mathbf{\Phi}}^*\underline{\mathbf{C}}_+^\dagger\mathbb{T}^\dagger \tag{94}$$

we regain equation (38) except for a change in the sign of the charge as required.

## 5.2 The Versatile Radiation Equation

We discuss the behaviour of the radiation equation under spatial and temporal rotation in Euclidean spacetime.

We first show that the current behaves like a Euclidean four-vector. Since we are leaving the components of the quaternions invariant and transforming the quaternion matrices, we must define $\mathbf{H}^{-1}(\mathbf{J})$ as the Euclidean four-vector current for transformational properties. We now set the properties of $\underline{\mathbf{I}}_\mu$ and $\underline{\mathbf{K}}_\mu$ under spatial rotation and Lorentz transformation.

$$\underline{\mathbf{I}}'_\mu = \mathbb{R}\,\underline{\mathbf{I}}_\mu\mathbb{R}^\ddagger, \qquad \underline{\mathbf{K}}'_\mu = \mathbb{R}^n\underline{\mathbf{K}}_\mu\mathbb{R}^{\ddagger n} \tag{95}$$

where $n$ is as described in section 5.1 and $\mathbb{R}$ is $\mathbb{R}_S$ for a spatial rotation and $\mathbb{R}_T$ for a temporal rotation. We may discover the rotational properties of $\underline{\mathbf{\Phi}}_S$ from its definition, equation (48), and the rotational properties of $\phi_1$ and $\phi_2$ given in equation (86). We have, for a spatial and temporal rotation, respectively,

$$\underline{\mathbf{\Phi}}'_S = \mathbf{R}_S|^n \underline{\mathbf{\Phi}}_S \mathbf{R}_S|^\ddagger, \qquad \underline{\mathbf{\Phi}}'_S = \mathbf{R}_T|^n \underline{\mathbf{\Phi}}_S \mathbf{R}_T|^\ddagger \tag{96}$$

Equation (49) becomes

$$\mathbf{J}_\mu|' = \underline{\mathbf{K}}'_\mu \underline{\mathbf{\Phi}}'_S \underline{\mathbf{I}}'_\mu \underline{\mathbf{\Phi}}'_S \tag{97}$$

and hence from equations (86), (95) and (96) we have for a spatial rotation using $\mathbf{R}_S|$ and temporal rotations using $\mathbf{R}_T|$

$$\mathbf{J}_\mu|' = \mathbf{R}_S|^n \mathbf{J}_\mu| \mathbf{R}_S|^{\ddagger n}, \qquad \mathbf{J}_\mu|' = \mathbf{R}_T|^n \mathbf{J}_\mu| \mathbf{R}_T|^{\ddagger n} \tag{98}$$

Neither transformation alters the temporal components of the quaternion elements of $\mathbf{J}_\mu|$. Hence

$$t:\left(\mathbf{J}_\mu|'\right) = t:\left(\mathbf{J}_\mu|\right) \tag{99}$$

and, from equation (50), $J_\mu^\S$ is invariant under spatial rotation and Lorentz transformation. This enables us to set the transformational properties of $\underline{\mathbf{J}}$ as defined in equation (54) as

$$\underline{\mathbf{J}}' = \mathbf{R}|\, \underline{\mathbf{J}}\, \mathbf{R}|^\ddagger \tag{100}$$

where, again, $\mathbf{R}|$ may be $\mathbf{R}_S|$ or $\mathbf{R}_T|$. The current, $J_\mu^\S$, is the same for both the original Dirac equation (19), and, immediately after its translation, the versatile Dirac equation (38). From section 4, we see that $\mathbf{H}^{-1}(\mathbf{J})$ behaves like a Euclidean four-vector, as does the usual current for the original Dirac equation, and hence these must remain equal in all frames.

Second, we show that the radiation equation, equation (53), is invariant. This follows immediately from equations (81), (82) and (100).

Third, we show that the Dirac current is conserved in all frames. Premultiplying equations (33) with zero potential by the Hermitian conjugates of $\phi_1$ and $\phi_2$, respectively, we obtain

$$\phi_1^\dagger\left(\mathbf{D}^\ddagger\phi_1\right) = \phi_1^\dagger\phi_2\mathbf{M}, \qquad \phi_2^\dagger\left(\mathbf{D}\phi_2\right) = -\phi_2^\dagger\phi_1\mathbf{M}^\ddagger \tag{101}$$

where the parentheses indicate the direction of differentiation. We form the Hermitian conjugates of equations (33) and postmultiply by $\phi_1$ and $\phi_2$, respectively, we obtain

$$\left(\phi_1^\dagger\mathbf{D}^\ddagger\right)\phi_1 = \mathbf{M}\phi_2^\dagger\phi_1, \qquad \left(\phi_2^\dagger\mathbf{D}\right)\phi_2 = -\mathbf{M}^\ddagger\phi_1^\dagger\phi_2 \tag{102}$$

Let

$$D_0 = i\,\partial/\partial x_0^{\sim}, \qquad D_r = \partial/\partial x_r \tag{103}$$

Adding equations (101) and (102) we obtain

$$\begin{aligned} &\sum_\mu \left[D_\mu\left\{\phi_1^\dagger\mathbf{i}_\mu^\ddagger\phi_1 + \phi_2^\dagger\mathbf{i}_\mu\phi_2\right\}\right] \\ &= \phi_1^\dagger\phi_2\mathbf{M} - \phi_2^\dagger\phi_1\mathbf{M}^\ddagger + \mathbf{M}\phi_2^\dagger\phi_1 - \mathbf{M}^\ddagger\phi_1^\dagger\phi_2 \end{aligned} \tag{104}$$

We multiply by $(-i/4)$ and put the final result in invariant form, using equations (48), obtaining

$$\sum_{\mu}\left[D_{\mu}\left(t\text{: Trace}\left\{\underline{\mathbf{K}}_{\mu}\underline{\mathbf{\Phi}}_{S}\underline{\mathbf{I}}_{\mu}\underline{\mathbf{\Phi}}\right\}\right)\right] \tag{105}$$
$$= t\text{: }\left(\text{Trace}\left\{\underline{\mathbf{K}}_{\mu}\underline{\mathbf{\Phi}}_{S}\underline{\mathbf{\Phi}}\underline{\mathbf{M}} + \underline{\mathbf{K}}_{\mu}\underline{\mathbf{M}}\underline{\mathbf{\Phi}}_{S}\underline{\mathbf{\Phi}}\right\}\right)$$

Spatial or temporal rotations of the expressions in curly brackets on both sides of this equation induce a similarity transformation. The expression in round brackets on the right hand side of equation (105) is zero when $\mathbf{M} = m$, and, since only similarity transformations are applied in changing frames, the right hand side of the equation remains zero in all frames. The temporal component of the quantity in round brackets on the left hand side of the equations equals the Dirac current, $J_{\mu}^{\S}$, from equations (49) and (50) and is constant from equation (99). Hence the Dirac current is conserved in all frames for all values of *n*.

In the above $D_{\mu}$ does not vary. If we wish to vary $D_{\mu}$ as a Euclidean four-vector, we must induce a similar variation for $J_{\mu}^{\S}$ by replacing it with $\mathbf{H}^{-1}(\mathbf{J})$. Then the left-hand side of equation (105) is equal to the dot products of Euclidean four-vectors and is hence constant and thus always zero.

## 6. DISCUSSION

We have derived a new version of the original Dirac and radiation equations of quantum electrodynamics using quaternions in Euclidean spacetime. The new version of the equations, called the versatile equations, is identical in every respect to the original equations if half-angle behaviour of the phase of the wave function is assumed for spatial and temporal rotations in Euclidean spacetime. That is, every solution of the original equations is, via a lift or map of the wave function, a solution of the versatile version with the same eigenvalues, and vice versa.

However, the versatile version allows more general behaviour of the wave function under spatial or temporal rotation. It does not have to be half angular. Other phase changes leave the equations invariant and the eigenvalues identical. One of the phase changes that is allowed has the versatile bispinor behaving as a Euclidean four-vector under spatial and temporal rotation in Euclidean spacetime. For such behaviour the versatile version of the equations may not be mapped back to the original version and hence the versatile versions constitute a generalisation of the original.

This behaviour under spatial rotation and Lorentz transformation has implications for the spin of the particle. The phase change, $\varphi$, is tied to the spin of the particle via the equation

$$\varphi = n\theta \tag{106}$$

where $\theta$ is the angle through which the particle is rotated about the quantisation axis and $n$ is the component of spin in this direction. This has implications for both the dynamical and Berry phase.(4) Experiments confirm that the phase changes for fermions obeying the Dirac equation are those expected for a spin of a half.(4) If $n = 1$ the spin of the particle must be integer. The model which used the Dirac and radiation equations for, say, $n = 1$ would have to take this into account.

A new method of calculating the spin of the particle from the Dirac equation might be required, but this need not necessarily involve quantum mechanics. Potentially, this generalisation to a value of $n = 1$ would permit the equations to be applied to a classical system in which the analogues of the bispinors were Euclidean four-vectors. No satisfactory Dirac current could be defined for previous attempts to permit such behaviour. We have found a way to provide the generalised equations with a current identical to Dirac's.

We could therefore summarise our most surprising finding by saying that, potentially, the Dirac and radiation equations could be applicable to classical problems with the bispinor defined to be, for example, the position vector of a classical particle. In order to do this one would require a special interpretation of the algebra and transformations which we will give examples of in our future work.

## ACKNOWLEDGEMENTS

One of us (Bell) would like to thank J. Dylan Morgan for alerting us to the importance of quaternion quantum mechanics and to acknowledge the assistance of E. A. E. Bell.